\documentclass[twocolumn, amsmath, amssymb, aps, physrev]{revtex4-2}
\usepackage{xcolor}
\usepackage{graphicx}
\usepackage{dcolumn}
\usepackage{bm}
\usepackage{physics}
\usepackage{braket}
\usepackage{soul}
\usepackage{ulem}
\begin{document}

\preprint{APS/123-QED}

\title{Scalable universal quantum gates between nitrogen-vacancy centers in levitated nanodiamond arrays}

\author{Guangyu Zhang}
\email{3120246074@bit.edu.cn}
\affiliation{Center for Quantum Technology Research and Key Laboratory of Advanced Optoelectronic Quantum Architecture and Measurements (MOE),
School of Physics, Beijing Institute of Technology, Beijing 100081, China}

\author{Huaijin Zhang}
\affiliation{Center for Quantum Technology Research and Key Laboratory of Advanced Optoelectronic Quantum Architecture and Measurements (MOE),
School of Physics, Beijing Institute of Technology, Beijing 100081, China}

\author{Zhangqi Yin}
\email{zqyin@bit.edu.cn}
\affiliation{Center for Quantum Technology Research and Key Laboratory of Advanced Optoelectronic Quantum Architecture and Measurements (MOE),
School of Physics, Beijing Institute of Technology, Beijing 100081, China}
\altaffiliation{Corresponding author} 
\date{\today}

\begin{abstract}
\quad Nitrogen-vacancy (NV) centers in nanodiamond offer a promising platform for quantum information processing due to their room-temperature spin coherence and optical addressability. However, scalable quantum processors remain limited by the challenge of achieving strong, controllable interactions between distant NV spins. Here, we propose a scalable architecture utilizing optically levitated nanodiamond arrays, where torsional vibrations mediate the coherent coupling between the embedded NV centers. By optimizing the shape of ellipsoidal nanoparticles, we achieve a light-induced coupling strength exceeding 119 kHz between torsional modes of the distant levitated nanodiamonds, which are two orders of magnitude larger than the typical decoherence rates in this system. This strong interaction, combined with magnetic field-enabled spin-torsion coupling, establishes an effective interaction between the spatially separated NV centers in the distant nanodiamonds. Numerical simulations confirm that dynamic decoupling can suppress both thermal noise and spin dephasing, enabling two-qubit gates with fidelity exceeding 99\%. This work provides a foundation for reconfigurable quantum hybrid systems, with potential applications in rotational sensing and programmable quantum processing. 
\end{abstract}

\maketitle

\section{INTRODUCTION}
\label{sec:intro}

The pursuit of scalable quantum processors has seen remarkable progress across multiple physical platforms. Superconducting circuits leverage mature fabrication techniques but require milli-Kelvin temperatures \cite{blais2021circuit}, while trapped ions offer exceptional coherence at the cost of slow gate speeds and complex vacuum systems \cite{haffner2008quantum}. Neutral atom arrays provide spatial reconfigurability \cite{henriet2020quantum}, but face fundamental limits in gate fidelity and optical addressing density. Among the solid-state candidates, nitrogen vacancy (NV) centers in diamond emerge as a uniquely promising platform, combining room temperature spin coherence \cite{bar2013solid}, nanoscale integration capacity \cite{millen2014nanoscale}, and individual optical addressability \cite{neumann2010single}. These inherent advantages position NV centers as prime candidates for scalable quantum technologies - provided that the critical challenge of establishing controllable long-range spin interactions can be overcome.

Existing approaches for NV-center-based quantum processing face critical trade-offs between scalability and control fidelity. Direct dipolar coupling schemes \cite{dolde2013room} suffer from rapidly diminishing interaction strengths ($\propto1/r^3$) and require ultra-strong  magnetic field gradients ($\sim$10$^4$ T/m) for multiqubit operations \cite{rabl2009strong}. Hybrid architectures employing photonic cavities \cite{schroder2017scalable,greentree2016nanodiamonds} or superconducting resonators \cite{zhu2011coherent} improve the interaction range at the cost of introducing heterogeneous interfaces that degrade spin coherence and complicate large-scale integration. Although phonon-mediated coupling through bulk mechanical resonators \cite{rabl2010quantum} theoretically enables long-range interaction, experimental realizations using fixed nanostructures lack the dynamical reconfigurability essential for error-corrected quantum processors.

This fundamental tension between interaction range and system control motivates our exploration of optically levitated nanodiamonds as a paradigm-shifting platform. Recent breakthroughs in ground-state cooling \cite{jain2016direct} and manipulation of torsional modes \cite{hoang2016torsional} have established levitated nanoparticles as pristine quantum oscillators with ultralow decoherence rates. Crucially, the anisotropic polarizability of nonspherical particles generates optical binding forces ($\sim$MHz) that couple torsional modes across nanoparticle arrays \cite{yan2023demand} - a mechanism two orders of magnitude stronger than phonon decoherence channels. Combined with the magnetic field-induced strong coupling between the torsional modes and the NV centers \cite{yin2013large}, this creates an unprecedented opportunity to mediate high-fidelity quantum gates between the distant qubits of the NV centers while preserving the addressability of the individual qubits.

In this study, we present a scalable quantum computation architecture based on optically levitated nanodiamond arrays with embedded NV centers. The architecture achieves a dipole-dipole torsional coupling strength of 119 kHz, facilitated by the design of ellipsoidal particles and the engineering of optical binding fields.  The implementation of dynamical decoupling enables fault-tolerant two-qubit gate fidelity of up to $99\%$. The scalability of this architecture is attributable to two primary features: wavelength-selective addressing of individual NV centers and reconfigurable arrays geometries enabled by optical tweezers networks. Furthermore, we demonstrate how these capabilities support the implementation of quantum error correction protocols and the development of modular quantum networks.

The structure of this paper is organized as follows: Section II explores the torsional coupling mechanism and analyzes the tunable coupling strength. The third section details the hybrid spin-torsion interaction model and the implementation of quantum gates. Finally, decoherence channels, scalability pathways, and potential applications are examined.

\section{TORSIONAL MODE COUPLING}
\label{sec:level2}

Although previous investigations of optically levitated nanoparticle arrays have focused on dipole-dipole interactions between center-of-mass (CoM) motion modes, the estimated coupling strengths are of the order of $\sim$100 Hz  \cite{liu2020prethermalization}. 
The effective coupling between two levitated nanoparticles can be enhanced to kilohertz through an optical cavity \cite{Vijayan2024Cavity}. However, this remains insufficient for high fidelity quantum information processing. 
This limitation motivates our exploration of torsional modes, whose fundamental frequencies typically exceed those of CoM modes by an order of magnitude \cite{hoang2016torsional,knowles2014observing}. The enhanced frequency scale facilitates quantum ground-state cooling for torsional degrees of freedom. In this section, a comprehensive analysis of the coupling mechanism between torsional modes in optically levitated nanodiamonds is presented, and parameter regimes for strong coupling are established.

\subsection{Dipole-Dipole Interaction Model}
\label{sec:coupling}
\begin{figure*}[htbp]
\centering
\includegraphics[width=0.8\textwidth]{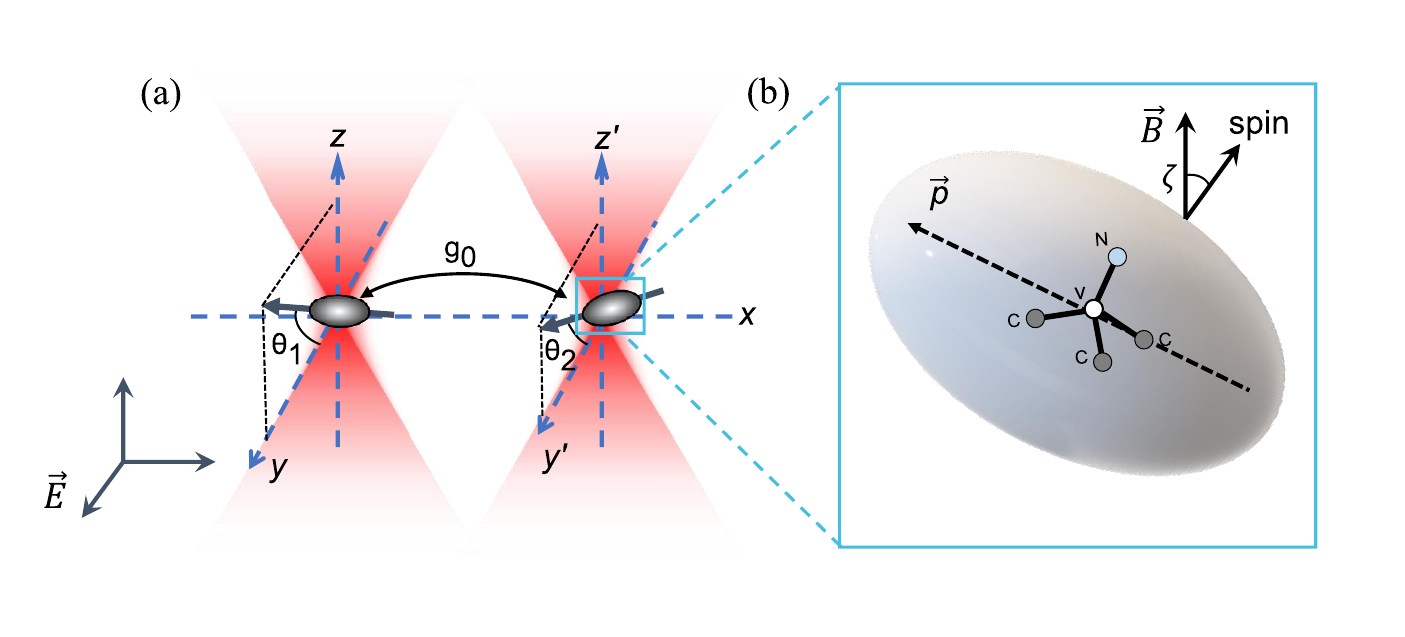} 
\caption{\label{Ellipsoid_array} 
(a) Experimental schematic: Two ellipsoidal nanodiamonds containing NV centers are optically trapped by  linearly polarized laser beams propagating along z-axis in high vacuum. Torsional motion occurs in the $y$-$z$ plane with angular deviations $\theta_{1,2}$ from the laser beams polarization direction along y-axis. 
The two torsional modes couple with each other under dipole-dipole interaction strength $g_0$.
(b) Under the external homogeneous magnetic field $\hat{B}$, the angle $\zeta$ between $\hat{B}$ and the axis of the NV center electron spin changes with the torsional motion, leading to the strong spin-torsion coupling \cite{Ma2017Proposal}.
}
\end{figure*}

Consider two ellipsoidal nanodiamonds trapped in linearly polarized optical tweezers propagating along z-axis under high vacuum, as depicted in Fig.\ref{Ellipsoid_array}(a). We assume that the long and short semiaxis of the nanodiamonds $a,b$ is much less than the wavelength $\lambda$ of the trapping lasers. Therefore, the dipole approximation can be applied here.  
Each particle exhibits anisotropic polarizability described by a diagonal tensor \cite{bohren2008absorption}:
\begin{equation}
\bm{\alpha} = \begin{pmatrix}
\alpha_{\parallel} & 0 & 0 \\
0 & \alpha_{\perp} & 0 \\
0 & 0 & \alpha_{\perp}
\end{pmatrix}, \quad \alpha_{\parallel} > \alpha_{\perp},
\end{equation}
where $\alpha_{\parallel}$ ($\alpha_{\perp}$) denotes the polarizability component parallel (perpendicular) to the trapping field. These components depend on the particle volume $V = 4\pi a b^2/3$, and depolarization factors $L_j$ through:
\begin{equation}
    \alpha_j = \frac{\epsilon_0(\epsilon_r - 1) V}{1 + L_j(\epsilon_r - 1)}.
\end{equation}
Here $\epsilon_0$ is the vacuum permittivity, $\epsilon_r$ is the relative permittivity of nanodiamond and $L_j$ is determined by the eccentricity $e = \sqrt{1 - (b/a)^2}$
\begin{equation}
    \begin{aligned}
        &L_{\parallel} = \frac{1-e^2}{e^3}\left(\frac{1}{2}\ln\frac{1+e}{1-e} - e\right), \\
        &L_{\perp} = \frac{1 - L_{\parallel}}{2}. 
    \end{aligned}
\end{equation}

Under dipole approximation, the dipole-dipole interaction potential for particles separated by distance $\mathbf{\hat{R}}$ along $x$-axis is \cite{dholakia2010colloquium,rieser2022tunable}:
\begin{align}
    U_{\text{int}} = \frac{e^{\mathrm{i}kR}}{4\pi\epsilon_0 R^3} \Biggl[ 
        &\Bigl[ \mathbf{p}_1 \cdot \mathbf{p}_2 
        - 3(\mathbf{p}_1 \cdot \hat{\mathbf{R}})(\mathbf{p}_2 \cdot \hat{\mathbf{R}}) \Bigr](1 - \mathrm{i}kR) \notag \\
        &- k^2R^2 \Bigl( (\mathbf{p}_1 \times \hat{\mathbf{R}}) \cdot (\mathbf{p}_2 \times \hat{\mathbf{R}}) \Bigr) \Biggr].
    \label{Uint}
\end{align}
Here, the wavenumber $k = 2\pi/\lambda$, $\mathbf{p}_i$ denotes the dipole moment of the $i$-th nanodiamond. With $y$-polarized trapping fields of central strength $E_0$, nanodiamonds align with $\mathbf{\hat{y}}$ to minimize potential energy. For small angular deviations $\theta_i \ll 1$, the $i$-th nanodiamond's dipole moment develops transverse components:
\begin{equation}
\mathbf{p}_i \approx \alpha_{\parallel} E_0 \mathbf{\hat{y}} + (\alpha_{\parallel} - \alpha_{\perp})E_0 \theta_i \mathbf{\hat{z}}.
\label{eq:p}
\end{equation}
Substituting Eq. (\ref{eq:p}) into Eq. (\ref{Uint}) and retaining the second-order term of $\theta_i$, the dipole-dipole interaction potential becomes
\begin{equation}
U_{\text{int}} \approx -(\alpha_{\parallel} - \alpha_{\perp})^2 E_0^2\text{Re}(G_{zz})\theta_1 \theta_2,
\end{equation}
where the component of the dyadic Green's function $G_{zz} = e^{ikR}[k^2R^2 + ikR - 1]/4\pi\epsilon_0 R^3$ characterizes the z-axis dipole-dipole interaction between $\mathbf{p}_1$ and $\mathbf{p}_2$, with the perpendicular alignment geometry inducing a restoring torque that maintains angular stability \cite{dholakia2010colloquium,rudolph2024quantum}.

Small angular displacements generate harmonic torsional potentials:
\begin{equation}
U(\theta_i) \approx \frac{1}{2}k_{\text{tor}}\theta_i^2.
\end{equation}
Here $k_{\text{tor}} = V(\chi_{\parallel} - \chi_{\perp})I_0/c$, $\chi_{\parallel,\perp} \equiv \alpha_{\parallel,\perp}/(\epsilon_0V)$ are normalized polarizabilities, $c$ is the speed of light and $I_0 = 2P_0/\pi w_0^2$ is the intensity of trapping laser, in which $P_0$ is the laser power and $w_0$ is the radius of beam waist.

If the motion of the nanodiamonds is cooled near the ground states, a Hamiltonian with quantized torsional modes is needed to describe the system. Quantizing the torsional modes through annihilation (creation) operators $b_i$ ($b_i^\dagger$) and neglecting the fast rotating terms, the system Hamiltonian reads:
\begin{equation}
H = \sum_{i=1,2}\hbar\omega_i b_i^\dagger b_i + \hbar g_0(b_1^\dagger b_2 + b_2^\dagger b_1),
\end{equation}
where the torsional frequency $\omega_i=\sqrt{k_{i,\text{tor}}/I_i}$ and coupling strength $g_0$ is given by:
\begin{equation}
    g_0 = \text{Re}(G_{zz})\frac{\Delta\alpha^2 E_0^2}{2\sqrt{I_1I_2\omega_1\omega_2}},
\label{eq:g0}
\end{equation}
where $\Delta\alpha=\alpha_{\parallel} - \alpha_{\perp}$ denotes the polarizability anisotropy, $I_i = 4\pi\rho ab^2(a^2 + b^2)/15$ is the moment of inertia, $\rho$ is the diamond density. In the far-field regime, where $kR$ is much larger than $1$ and $g_0\sim R^{-1} $, the coupling strength decays much more slowly than the Coulomb force. This behavior enables both strong short-range coupling and extended interaction ranges compared to direct Coulomb interactions. For parameters $a = 300$ nm, $b = 180$ nm, $R = 1.06\ \mu\mathrm{m}$, $\lambda = 1064$ nm, we can estimate  $g_0/2\pi = 119$ kHz, which exceeds typical environmental decoherence rates by two orders of magnitude \cite{knowles2014observing}. In this way, the levitated nanodiamonds array with torsional modes coupling becomes a promising platform for quantum information processing.

\subsection{Tunable coupling strength}
\label{sec:level2}
\begin{figure}[htbp]
\centering
\includegraphics[width=0.5\textwidth]{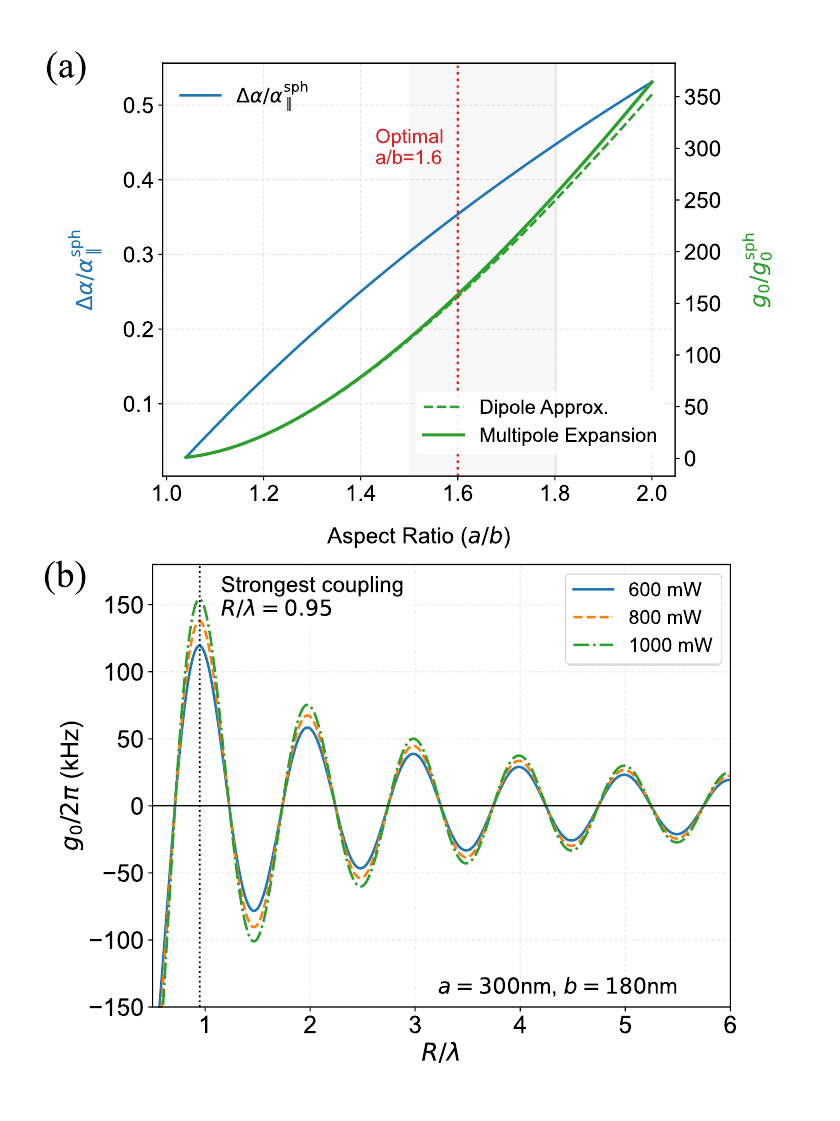}
\caption{\label{g(e,R)} 
(a) Normalized coupling strength vs. particle aspect ratio. The normalized coupling strength $g_0/g_0^{\text{sph}}$ (right axis, green lines) and polarizability anisotropy $\Delta\alpha/\alpha_{\parallel}^{\text{sph}}$ (left axis, blue line) are plotted against the ratio $a/b$. Here, $g_0^{\text{sph}}$ denotes the coupling strength of a spherical particle ($a/b=1$, $a=250$ nm) with identical volume, and $\alpha_{\parallel}^{\text{sph}}$ is the polarizability of the spherical particle along its symmetry axis. The green dashed curve incorporates multipole corrections (quadrupole and higher-order terms), revealing a optimal point at $a/b=1.6$ (vertical dotted line) compared to the spherical case. (b) Tunable coupling strength via interparticle spacing at optimal ratio. The coupling rate $g_0/2\pi$ is modulated by varying the normalized spacing $R/\lambda$ for three laser powers (600, 800, 1000 mW) using particles ratio with $a/b=1.6$. The strongest coupling occurs at $R/\lambda=0.95$ (vertical dashed line), where near-field interactions dominate. All calculations assume fixed particle volume($a=300$ nm, $b=180$  nm).
}
\end{figure}

As Eq.\eqref{eq:g0} illustrated, the coupling strength depends on three key parameters: particle size, interparticle spacing, and laser power. The particle size is fixed during sample preparation, while the latter two parameters can be dynamically adjusted by tuning the optical tweezers' intensity and the interparticle distance.

Although the dipole approximation in Eq. \eqref{eq:g0} provides the theoretical basis for the coupling strength, and higher-order multipole effects become significant when the dimensions of the particle approach the optical wavelengths (\(a > \lambda/4\)). Fig. \ref{g(e,R)}(a) reveals a notable deviation between the dipole-approximated results (green solid curve) and the multipole-corrected solutions (green dash curve) for aspect ratios $a/b > 1.4$. The observed discrepancy can be attributed to the presence of a localized field enhancement at the particle tips. The high curvature present at the ellipsoid extremities serves to amplify the charge density gradients, thus inducing strong quadrupole moments \cite{pesce2020optical}. Quantitative corrections for these effects are derived in Appendix \ref{app:coefficient_calculation}.

The optimal ratio \(a/b=1.6\) represents a compromise between two key considerations: (1) the saturation of \(\Delta\alpha\) growth, evidenced by the reduced slope of the blue curve for \(a/b>1.8\) in Fig.\ref{g(e,R)}(a); and (2) thermal noise suppression requirements (\(g_0 > 10\Gamma_{\mathrm{th}}\)). Sensitivity analysis reveals coupling fluctuations below \(1\%\) within \(a/b=1.6\pm0.1\), indicating robust tolerance to fabrication variations.

The power-dependent tunability is shown in Fig. \ref{g(e,R)}(b), which results from the optical trap field modulation. Increasing laser power from 600 mW to 1000 mW enhances the electric field strength \(E_0 \propto \sqrt{P_0}\), thereby increasing the polarizability anisotropy \(\Delta\alpha\) and amplifying the coupling strength \(g_0 \propto E_0^2\). Near-field effects dominate at \(R/\lambda=0.95\) (vertical dashed line), yielding a peak coupling strength of \(g_0= 2\pi \times 150\)~kHz. A full-width-at-half-maximum analysis ($\sim$0.2\(\lambda\)) further shows that stronger fields (1000 mW) extend the effective coupling range to \(R=1.2\lambda\), providing flexibility for dynamic arrays reconfiguration. In our model, we focus on symmetric lasers without relative phase, particularly at integer-wavelength particle spacing corresponding to coupling extrema , so recoil-scattering decoherence can be neglected\cite{PhysRevLett.133.233603}.

The central light intensity $I_0$ can be calculated through $I_0=E_0^2/2\eta_0$, where $E_0$ is the electric field amplitude, $\eta_0=1/\epsilon_0 c$ is the impedance of free space. The laser power reads $P_0=\pi I_0 \omega_0^2/2$, where $\omega_0$ is the beam waist radius. In this way, we can find that $E_0=\sqrt{4P_0/\pi\omega_0^2\epsilon_0 c}$. The simulation parameters are chosen as follows. The nanodiamond density $\rho=3500\text{ kg/m}^3$, relative permittivity $\varepsilon_r = 5.7$ at the operational frequency, wavelength $\lambda_0=1064\text{ nm}$, and beam waist radius $\omega_0=500 \text{ nm}$. 

Simulation results indicate that the maximum coupling strength between nanodiamonds significantly exceeds both the torsional mode rethermalisation rate $\sim100\text{ Hz}$ \cite{hoang2016torsional} and the NV center electron spin decay rate $\sim\text{ kHz}$ \cite{knowles2014observing}. The strong coupling conditions have been fulfilled, making the system particularly suitable for quantum information processing. As we will see in the next section, the gate speed is determined by the effective coupling strength.

\section{UNIVERSAL GATEs}
In order to realise quantum computation using NV centres as qubits, a major challenge is to achieve an effective coupling strength that exceeds the decay rates discussed in the last section. In this section we will show how to construct a set of universal quantum gates between the NV centres in the distant nanodiamonds.

\subsection{CPHASE gate protocol}
In our proposed model, two NV centers are embedded in distant nanodiamonds, oriented along different axes to induce different magnetic field shifts, allowing single qubit addressing via frequency domain. As discussed in Ref. \cite{chen2019universal}, the Hamiltonian for a single NV spin is given by $H_{NV} = DS_z^2 + \gamma\Vec{B}\cdot\Vec{S}$, where $D = 2.87 \text{ GHz}$ is the zero-field splitting \cite{maclaurin2012measurable}, $\gamma$ represents the gyromagnetic ratio and $\Vec{S}$ is the spin operator for the NV center electron spin. The torsional motion alters the relative angle $\zeta$ between $\Vec{B}$ and $\Vec{S}$, as shown in Fig. \ref{Ellipsoid_array}(b).

The effective Hamiltonian for this system, which includes the two NV centers and the torsional oscillators, can be expressed as follows:
\begin{equation}
    H = \sum_{i=1,2}[\omega_i b^{\dag}_i b_i + \frac{E_i}{2}\sigma_i^z + g_i\sigma_i^z (b_i^{\dag} + b_i)] + g_0 (b^{\dag}_1 b_2 + b^{\dag}_2 b_1 )
\label{eq:ham} 
\end{equation}
in which, $\omega_i$ represents the frequency of the torsional mode, where $E_i$ (for $i=1,2$) denotes the energy splitting between the eigenstates $\ket{S_z=-1}_i=\ket{1}_i$ and $\ket{S_z=0}_i=\ket{0}_i$ of spin-i in the two NV centers, determined by their relative angle to the magnetic field direction. The operator $\sigma_i^z$ is the $i$-th electron spin Pauli operator in the NV center $i$, and the coupling strength between the torsional mode and the $i$-th electron spin is $g_i=\sqrt{\frac{1}{8I\omega_i}}\frac{\partial E_i}{\partial\zeta}|_{\zeta=\zeta_i}$  \cite{Ma2017Proposal}. In order to maximize the spin-torsional coupling strength, we can control the torsional motion direction to be perpendicular to the two orientations of the NV centers and adjust the magnetic field to be parallel to the torsional motion plane.

In order to realize universal quantum computation, we need to construct two single-qubit gates and a controlled-phase (CPHASE) gate based on Eq. \eqref{eq:ham}. Single-qubit operations can be performed by adding on resonance drive along the x axis with Rabi frequency  $\Omega_i$ to induce a transition between states $\ket{0}$ and $\ket{1}$. When the Rabi frequency $\Omega_i$ is much larger than both coupling strength $g_i$ and $g_0$, in the interaction picture the single qubit gate Hamiltonian with the driving term is given by
\begin{equation}
    H_{s,i} = \frac{\Omega_i}{2}\sigma_i^x +g_i (b_i^{\dag} + b_i)\sigma_i^z
\end{equation}
where the spin-torsion coupling serves as an extra dephasing channel , with DD scheme based on Ref. \cite{rong2014implementation} the spin-torsion coupling-induced dephasing effect can be significantly reduced. 

In order achieve site-selective microwave addressing, we assumed that the distinct orientations of the two NV center electron spins relative to the applied magnetic field. This misalignment induces energy-level differences ($E_1 \neq E_2$), enabling frequency-resolved spin control and suppressing unintended spin driving. 

The CPHASE gate can be realized following the celebrated S${\o}$rensen-M${\o}$lmer gates scheme \cite{molmer1999multiparticle,sorensen2000entanglement}. In the interaction picture of $H_0=\frac{E_1}{2}\sigma_1^z + \frac{E_2}{2}\sigma_2^z$, we obtain the interaction Hamiltonian $H'$ and suppose that the torsional frequencies of two nanodiamonds are in resonance, $\omega_1 = \omega_2 = \omega_0$. Using a similar method as in Ref. \cite{chen2019universal}, the Hamiltonian $H'$ can be rewritten as
\begin{equation}
\begin{split}
    H' &= \sum_{i=1,2} \left[ \omega_i b_i^\dagger b_i + g_i \sigma_i^z (b_i^\dagger + b_i) \right] + g_0 (b_1^\dagger b_2 + b_2^\dagger b_1 ) \\
    &= U \left[ \Omega_1 b_1^\dagger b_1 + \Omega_2 b_2^\dagger b_2 - \left( \frac{S_1^2}{\Omega_1} + \frac{S_2^2}{\Omega_2} \right) \right] U^\dagger
\end{split}
\label{eq:H'}
\end{equation}
where $U$ is an unitary transformation 
\begin{equation}
    U=e^{b_1b_2^{\dag} - b_1^{\dag}b_2}e^{\frac{S_1}{\Omega_1}(b_1 - b_1^{\dag})}e^{\frac{S_2}{\Omega_2}(b_2 - b_2^{\dag})}
\end{equation}
in which $\Omega_1=\frac{1}{2}(\omega_0+g_0)$, and $\Omega_2=\frac{1}{2}(\omega_0-g_0)$, $S_1=(g_1\sigma_1^z+g_2\sigma_2^z)/2$, and $S_2=(g_2\sigma_2^z - g_1\sigma_1^z)/2$. The time evolution governed by the Eq. \eqref{eq:H'} reads
\begin{equation}
    e^{-iH't}=Ue^{-i(\Omega_1b^{\dag}_1 b_1 + \Omega_2b^{\dag}_2 b_2)t}e^{i\left( \frac{S_1^2}{\Omega_1} + \frac{S_2^2}{\Omega_2} \right)t}U^{\dag}
    \label{evolution}
\end{equation}
For certain times $t_m=2\pi m/(\Omega_1+\Omega_2)$ with $m=1,2,3\cdots$, the first exponential in Eq.\eqref{evolution} $e^{-i(\Omega_1b^{\dag}_1 b_1 + \Omega_2b^{\dag}_2 b_2)t}=I$. Here we suppose that the number operator $b^{\dag}_1 b_1$ and $b^{\dag}_2 b_2$ has the same integer. Thus Eq. \eqref{evolution} reduces to:
\begin{equation}
    e^{-iH't}=\text{exp}\left[\frac{2i\pi m}{\Omega_1+\Omega_2}\left( \frac{S_1^2}{\Omega_1} + \frac{S_2^2}{\Omega_2} \right)\right]
    \label{evolution2}
\end{equation}

Substituting $S_i^2 = g_1^2+g_2^2 \pm 2g_1g_2\sigma_1^z\otimes\sigma_2^z$ in Eq. \eqref{evolution2}, and ignoring the unimportant global phase we can get:
\begin{equation}
    e^{-iH't_m}=\text{exp}\left[i\frac{\pi m g_0g_1g_2}{2\omega_0\Omega_1\Omega_2}\sigma_1^z\otimes\sigma_2^z\right]
    \label{evolution3}
\end{equation}

The complete time evolution operator is expressed as $ e^{-iH't_m}e^{-iH_0t_m}$ in the laboratory reference frame. Using the spin echo technique, we can effectively eliminated the term $e^{-iH_0t_m} $. We first define a global flip operator $U_{\alpha}(\varphi)=\text{exp}(-i\varphi/2\sum_i\sigma_i^{\alpha})$  that induces a rotation of all qubits around the axis $ \alpha = x,y,z $. Using the basis $ \left\{ \ket{00},\ket{01},\ket{10},\ket{11} \right\} $ , substituting Eq. \eqref{evolution3} in the complete evolution under the spin echo technique :
\begin{equation}
    \begin{aligned}
        U(2t_m)&= U_x(\pi)e^{-iH't_m}e^{-iH_0t_m} U_x(\pi)e^{-iH't_m}e^{-iH_0t_m} \\  
        &= \text{diag}(e^{i\phi},1,1,e^{i\phi})      
    \end{aligned}
\end{equation}
where $\phi=\dfrac{8g_0g_1g_2m\pi}{\omega_0(\omega_0+g_0)(\omega_0-g_0)} $. By complementing the propagator $ U(2t_m) $ with $ U_z(-\phi) $, we obtain the CPHASE gate operation
\begin{equation}
    U_{CPHASE}=U_z(-\phi)U(2t_m)=\text{diag}(1,1,1,e^{2i\phi})  
\end{equation}

The gate time is $t_g=2t_m=4\pi m/\omega_0$. In order to achieve a controlled-Z gate $\phi=\pi/2$, the condition
\begin{equation}
    g_1g_2=\frac{\omega_0(\omega_0+g_0)(\omega_0-g_0)}{16g_0m}
\label{eq:gg} 
\end{equation}
must be fulfilled. Experimentally, the condition can be satisfied by tuning the optical tweezers power $P_0$ and waist $\omega_0$ (via $E_0\propto\sqrt{P_0}/\omega_0$), or modulating the external magnetic field $B$ to adjust the spin-torsion coupling $g_i$. For example, with $P_0=600$ mW and $B=0.1$ T, the required $g_i\approx2\pi\times350$ KHz is achievable within current levitation technologies \cite{Ma2017Proposal,delord2017strong,vijayan2023scalable}. The parameter $m$ must be an integer and its selection governs both the required coupling strengths $g_i$ and the achievable gate time $t_g$.

\subsection{Fidelity analysis}
\label{sec:level3b}
\begin{figure}[htbp]
\centering
\includegraphics[width=0.5\textwidth]{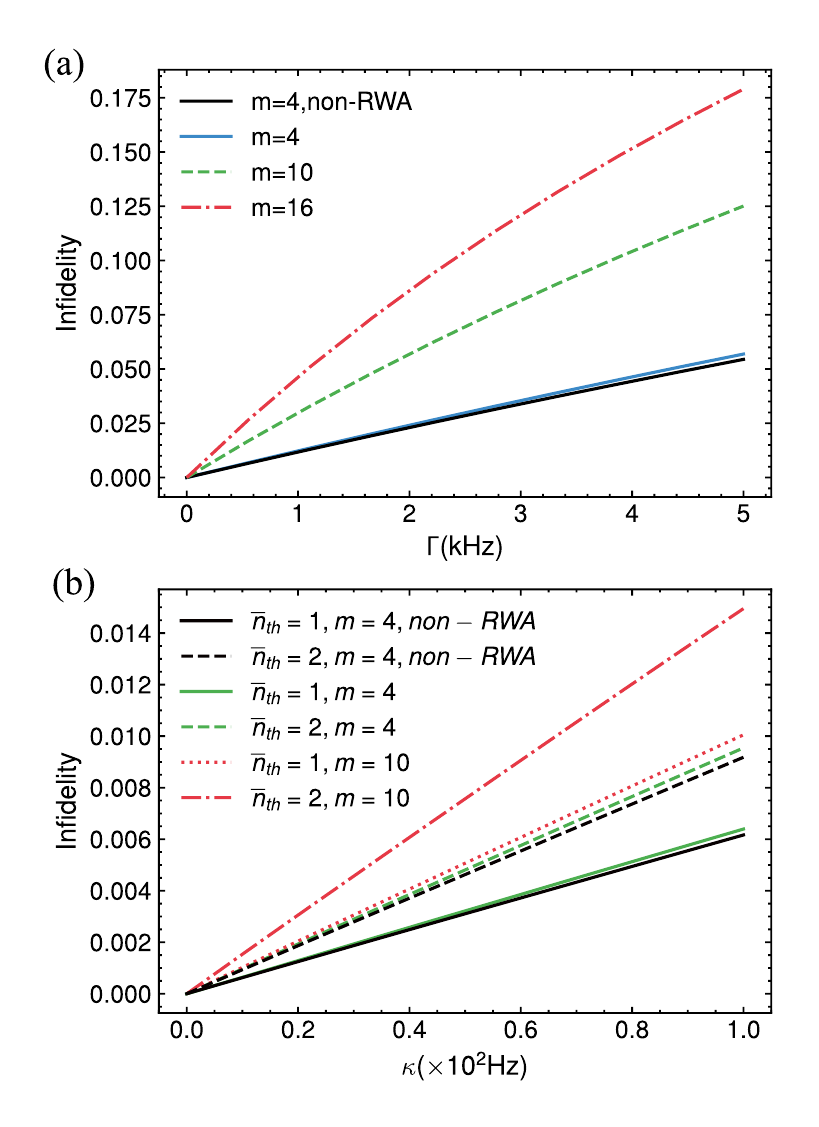}
\caption{\label{Infidelity_p} 
Infidelity due to qubit dephasing and rethermalization of the nanodiamond , with simulated both with and without(black curves) the rotating-wave approximation. (a) Dephasing of NV centers induced error for $m=4,10,16$, here $\kappa/\omega_i = 10^{-6}$ and $\overline{n}_{th}=0.01$. (b) Rethermalization of torsional mode induced errors for $\overline{n}_{th}=1$ and $\overline{n}_{th}=2$ with $m=4$ (green line) and $m=10$ (red line); here $\Gamma=0$.}

\end{figure}
In order to realize fault-tolerant quantum computation, achieving high-fidelity quantum gates is critical. Here we analyze the main decoherence sources in the hybrid system and evaluates their impact on gate fidelity. The dissipative dynamics of the system under decoherence are described by the master equation:
\begin{multline}
    \dot{\rho}=-i[H,\rho]+\kappa(\overline{n}_{th} + 1)\sum_{i=1,2}\mathcal{D}[b_i]\rho\\
    +\kappa\overline{n}_{th}\sum_{i=1,2}\mathcal{D}[b_i^{\dag}]\rho+\frac{\Gamma}{4}\sum_{i=1,2}\mathcal{D}[\sigma_i^z]\rho 
    \label{master}
\end{multline}
where $\mathcal{D}[b_i]\rho=b_i \rho b_i^{\dag}-\frac{1}{2}\left\{b_i^{\dag}b_i,\rho \right\}$. The second and third terms describe the rethermalization of the torsional mode to the thermal equilibrium state $\overline{n}_{\text{th}} = [e^{\hbar\omega_i/k_B T} - 1]^{-1}$, where $k_B$ is the Boltzmann constant and $T$ is the temperature. The decay rate of the torsional mode is given by $\kappa=\omega/Q$ \cite{gieseler2013thermal}, where $Q$ is the quality factor of the torsional mode. Highly pure nanodiamonds are essential to mitigate the heating effects of optical levitation \cite{frangeskou2018pure}. The last term in Eq.\eqref{master} describes the decoherence process of the NV centers with a decoherence rate $\Gamma\sim 1/T_2$, with $T_2$ as the spin coherence time. Here, assume that $T_1$ is typically much longer than $T_2$ by neglecting single-spin relaxation processes.

For the CPHASE gate, the initial product state is set as $\rho(0)=(\ket{00}+\ket{01}+\ket{10}+\ket{11})(\bra{00}+\bra{01}+\bra{10}+\bra{11})/4\otimes\rho_{th}(T)$. The fidelity is defined as $\mathcal{F}=\bra{\Phi_{tar}}\rho_{NV}\ket{\Phi_{tar}}$, where $\ket{\Phi_{tar}}=(\ket{00}+\ket{01}+\ket{10}-\ket{11} )/2$ and $\rho_{NV}=tr_a[\rho]$ is the density matrix of the NV centers, with $tr_a[\dots]$ the trace over the torsional mode degree of freedom. The simulation yields the infidelity $\xi=1-\mathcal{F}_{max}$, isolating contributions from rethermalization and dephasing for analysis, illustrated in Fig.\ref{Infidelity_p}.

The integer parameter \( m \), critical to determine the duration of the gate \( t_g = 4\pi m/\omega_0 \) is constrained by several interdependent physical and experimental conditions. To achieve the target phase \( \phi = \pi/2 \), $m$ must satisfy:
\begin{equation}
    m = \frac{\omega_0 (\omega_0 + g_0)(\omega_0 - g_0)}{16 g_0 g_1 g_2}.
\end{equation}

In experimentally feasible ranges, the gate time must satisfy $t_g<T_2/5$, where $T_2\sim 1$ ms, to minimize decoherence effects. Furthermore, condition $g_i<\omega_0$ prevents the hybridization of the mechanical and spin modes, with the upper bound of $g_i\leq 2\pi\times 500$ kHz imposed by the magnetic field strengths $B\leq0.1$ T and nanodiamond dimensions \cite{chen2019universal}. Meanwhile, $m$ must be integer-valued to ensure complete phase evolution in Eq. \eqref{eq:gg}. For a representative system with $\omega_0= 2\pi\times 1$ MHz and $g_0= 2\pi\times 100$ kHz, assuming $g_1=g_2$, we find $m\in[4,16]$. This corresponds to gate times of $t_g=22.5,56.3,90$ $\mu\text{s}$ for $m=4,10,16$, respectively, with spin-torsion coupling strengths $g_i=2\pi\times 460,290,250\text{ kHz}$. This parameter range maintains compatibility with fault-tolerance thresholds. For systems with limited $T_2$, dynamic decoupling protocols further suppress decoherence at higher $m$.

As demonstrated in Fig. \ref{Infidelity_p}, reducing $m$ improves gate fidelity but requires stronger spin-torsion coupling. Careful selection of $m$ is necessary to ensure the coupling strength remains below the torsional mode frequency. For $m=4$, the effective qubit dephasing rate of approximately $50\text{ Hz}$ leads to an infidelity of $\xi\approx0.5\%$. Under rethermalization conditions, where the effective dephasing reaches $\sim1$ kHz, the infidelity increases to $\xi\approx1\%$. Under the given parameters, $g_0/\omega_0\approx 0.1$, warranting consideration of the effects of the rotating wave approximation (RWA). Numerical integration of the full two-mode Hamiltonian, including the counter-rotating terms ($b_1^\dagger b^\dagger_2 + b_1 b_2 $), confirms that the gate infidelity is only marginally impacted, as illustrated by the black curves in Fig. \ref{Infidelity_p}.

The results suggest that the proposed optomechanical architecture can facilitate CPHASE gate fidelity exceeding $99$\%, outperforming conventional NV center implementations. Traditional magnetic dipole-dipole coupling schemes are limited to $\mathcal{F} < 95\%$ due to the $1/r^3$ interaction decay and ultrahigh magnetic gradient requirements ($\sim 10^4\ \mathrm{T/m}$) \cite{zhang2017selective}. While photonic cavity-mediated gates improve fidelity to $\sim 98\%$ \cite{greentree2016nanodiamonds}, they face challenges such as cryogenic operation constraints and cavity mode losses. Recent dynamical-decoupling-optimized protocols achieve CNOT gate fidelities up to $99.92\%$ using shaped microwave pulses \cite{xie202399}, but these methods rely on complex noise modeling and lack intrinsic scalability. By combining dynamical error correction techniques (e.g., pulse shaping) with phonon-mediated coupling, fidelities beyond $99\%$ could be attained while preserving the scalability of optical tweezer arrays. This hybrid approach meets the fault-tolerance requirements for surface codes ($F > 99\%$) \cite{gidney2021factor,xu2024constant,aasen2025topologically} and establishes levitated nanodiamonds as a promising platform for distributed quantum computing.

\subsection{Scalability analysis}
\begin{figure}[htbp]
\centering
\includegraphics[width=0.48\textwidth]{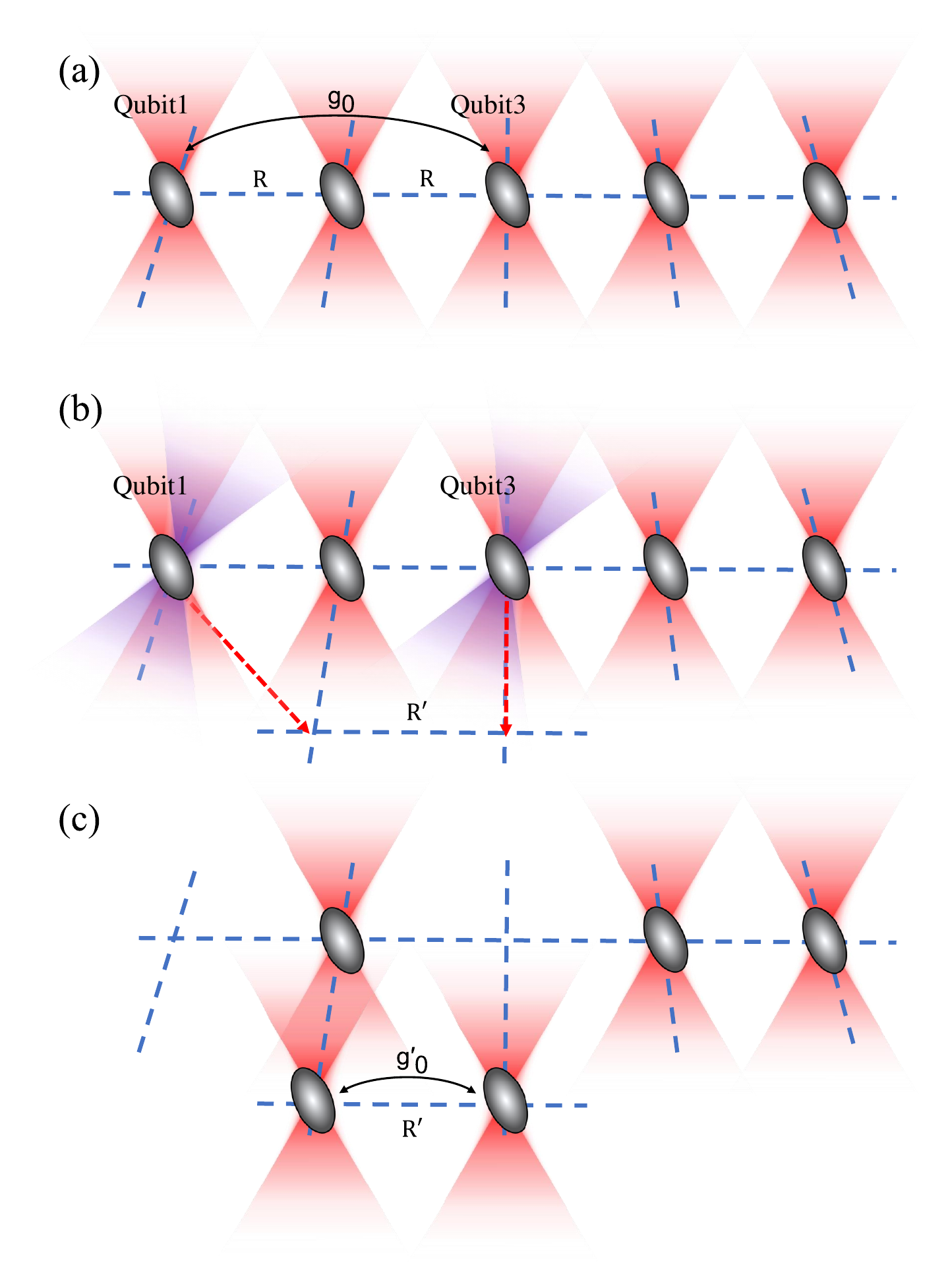}
\caption{\label{scalable} 
Diagram of reconfigurable levitated optomechanical arrays with NV centers. (a) In the initial arrays, nanodiamonds are trapped in linearly polarized optical tweezers under high vacuum. The torsional frequency is regulated by adjusting the laser power and beam radius. Nanodiamonds are coupled via dipole-dipole interactions, with each nanodiamond embedding an NV center that serves as the qubit carrier. (b) Target nanodiamonds (Qubit 1 and Qubit 3) are dynamically repositioned to non-local locations (indicated by red dashed lines) by modulating the wavelength, power, and beam radius of auxiliary optical tweezers (purple beams). This process forms a new effective coupling channel. (c) The dynamically reconfigured arrays supports non-local quantum gate operations, where the reduced inter-particle spacing $R'$ enhances the coupling strength $g'_0$, enabling high-fidelity CPHASE gate implementation. Coupling channels for the remaining unmanipulated particles can be deactivated via frequency detuning.
}
\end{figure}

Traditional coupling mechanisms, including direct dipole interactions, hybrid architectures employing photonic cavities or superconducting resonators, and phonon-mediated coupling via bulk mechanical resonators, face significant experimental challenges. These limitations include stringent operational requirements and restricted dynamical reconfigurability, both of which are critical for the implementation of error-corrected quantum processors.

In our proposed platform, the dipole-dipole coupling strength follows a scaling relation of \( g_0 \propto P_0/R \), offering superior long-range interaction capabilities compared to the \( 1/R^3 \) dependence of direct dipole coupling. Additionally, the uniform magnetic field required for spin-torsional coupling remains fully compatible with optical tweezers operation. As illustrated in Fig. \ref{g(e,R)}(b), this approach achieves a substantial coupling strength of \( g_0/2\pi \sim 50 \) kHz at a practical interparticle distance of \( R=2 \, \mu \mathrm{m} \), significantly surpassing typical thermal noise levels (\(\sim 100\) Hz) and enabling efficient nonlocal quantum operations.%

As discussed in Sec. \ref{sec:level2}, while the residual positioning error can be suppressed by tuning the power of optical tweezers (\( P_0 \)), simultaneously amplifies the recoil-induced decoherence from photon scattering\cite{PhysRevLett.133.233603}. However achieving sub-wavelength spacing remains experimentally challenging due to diffraction limits and inter-trap interference, an alternative theoretical approach involves utilizing ion-trap arrays to confine nanoparticles, followed by laser-induced coupling. This method reduces the required laser power, thereby mitigating particle rethermalization while enhancing dipole-dipole coupling strength, thereby relaxing the stringent subwavelength spacing constraints.

As discussed in Sec.\ref{sec:level3b} , this architecture supports high-fidelity quantum logic gate. Its strong compatibility with modular quantum computing allows for seamless integration with dynamic error correction techniques, such as shaped pulse optimization, achieving gate fidelities exceeding \( 99.9\% \) while maintaining scalability\cite{xie202399,xu2024constant}. The spectrally tunable coupling enables CPHASE gate implementation through resonant operation while suppressing off-resonant information exchange, which is a key advantage for scalable fault-tolerant quantum computing architectures.

The integration of reconfigurable optical tweezer arrays and tunable dipole-dipole couplings establishes a unified platform that overcomes previously scalability barriers in conventional NV-center based quantum computing architectures. Moreover, recent work on optical tweezer arrays has demonstrated precise placement ($\sim$10 nm) \cite{yan2023demand} of trapped particles, suggesting that large, ordered arrays of nanodiamonds could be assembled. This flexibility in layout and transport is functionally analogous to atom rearrangement in neutral atom systems\cite{jaksch2004optical} as depicted in Fig. \ref{scalable}, yet it is implemented here at room temperature and within a solid-state platform.

It is noteworthy that in experiments, fast modulation of optical tweezers using acousto-optic modulators (AOMs) can effectively turn off the dipole–dipole interaction by detuning the trapping light frequency far off-resonance with the polarizability spectrum of the induced dipole, that is, making the optical field off resonance with respect to the dipole response. The response time of this modulation is significantly shorter than that of the gate duration \cite{zheng2019cooling}, resulting in a minimal impact on the fidelity of the entangled state stored in the NV centers within the distant nanodiamonds. This adaptability addresses the dynamic coupling and topological requirements essential for large-scale quantum processors.

Recent advances, including ground-state cooling (\(\overline{n}_{th}<0.1\)) and all-optical feedback cooling \cite{vijayan2023scalable, fonseca2016nonlinear}, have significantly suppressed thermal noise in large-scale arrays. Combined with recent progress in scalable nanodiamond fabrication \cite{shulevitz2022template} and low-overhead qLDPC codes \cite{xu2024constant}, fault-tolerant quantum computation appears increasingly feasible using arrays of a few thousand well-characterized NV qubits.

Although our system does require initial characterization per qubit - unlike neutral atoms that are intrinsically identical - it offers complementary strengths: long NV spin coherence times, high optical stability and strongly controllable long-range interactions. These features position optically levitated nanodiamond arrays as a promising alternative pathway toward fault-tolerant quantum computing with reduced overhead and ambient-condition operation.

Extending this model to multiparticle configurations facilitates investigations of macroscopic quantum phenomena, analogous to studies conducted in ultra-cold atom systems \cite{kaufman2021quantum,bernien2017probing}. Potential applications include the measurement of entanglement entropy, the exploration of quantum phase transitions \cite{keesling2019quantum}, and the characterization of quantum many-body scars \cite{serbyn2021quantum}. Precise spatial control of nanodiamonds, achieved through stochastic loading and ultrasonic atomization \cite{vijayan2023scalable}, is essential for the assembly of predefined trapping arrays. Moreover, optical manipulation enables non-local gate operations and distributed quantum architectures, establishing a versatile platform for scalable quantum computation and the study of complex quantum phenomena \cite{Zhang2021}.

\section{CONCLUSION}
\quad In conclusion, we have introduced a novel mechanism to mediate interactions between spatially separated NV centers via torsional motion in distant optically levitated nanodiamonds. Our approach achieves a torsion-torsion coupling strength of approximately 119 kHz, significantly exceeding the NV spin dephasing rate and torsional mode decay rate by two orders of magnitude. The system's tunable coupling parameters and precise optical control make it well-suited for exploring collective quantum phenomena and scalable many-body interactions in hybrid quantum systems. This framework supports high-fidelity CPHASE gate operations with fidelity exceeding $99\%$, meeting the thresholds required for fault-tolerant quantum computation. Furthermore, the integration of programmable dipole-dipole interactions via optical tweezers arrays overcomes scalability limitations inherent in conventional NV-center-based architectures. Beyond high-fidelity quantum gates, this platform has strong potential for quantum sensing, and programmable quantum simulation.

\begin{acknowledgments}
This work is supported by the National Natural Science Foundation of China (Grant No. 12441502), the Beijing Institute of Technology Research Fund Program for Young Scholars, and Tianyan Quantum Computing Program. 
We thank Xingyan Chen and Shengyan Liu for helpful discussions.

The data that support the findings of this article are openly available \cite{zhang_2025_16917848}.
\end{acknowledgments}

\appendix
\section{Multipole Expansion Coefficients for Coupling Strength}
\label{app:coefficient_calculation}
The electrostatic interaction energy between two ellipsoidal particles, expressed as a multipole series:
\begin{equation}
U_{\text{multipole}} = \frac{1}{4\pi\epsilon_0} \left[ \frac{\mathbf{p}_1 \cdot \mathbf{p}_2}{R^3} + \frac{\mathbf{p}_1 \cdot \mathbf{Q}_2 \cdot \mathbf{R}}{R^5} + \cdots \right],
\end{equation}
where \(\mathbf{p}_i\) and \(\mathbf{Q}_i\) denote dipole and quadrupole moments, respectively. The quantum optomechanical coupling strength \(g_0\) is then proportional to the interaction energy, leading to:
\begin{equation}
g_0^{\text{exact}} = g_0^{\text{dip}} \left( 1 + \sum_{n=1}^\infty C_n \left(\frac{a}{\lambda}\right)^n \right),
\end{equation}
where \(g_0^{\text{dip}}\) represents the dipole-approximated coupling strength. Truncating to second order, we obtain:
\begin{equation}
\frac{g_0^{\text{exact}}}{g_0^{\text{dip}}} = 1 + C_1 \left(\frac{a}{\lambda}\right) + C_2 \left(\frac{a}{\lambda}\right)^2 + \mathcal{O}\left(\frac{a}{\lambda}\right)^3.
\end{equation}

To resolve the coefficients \(C_1\) and \(C_2\), we use prolate spheroidal coordinates \((\xi, \eta, \phi)\). The electric potential is expanded as\cite{landau2013electrodynamics}:
\begin{equation}
\Phi(\xi) = \sum_{n=0}^\infty A_n Q_n(\xi) P_n(\eta),
\end{equation}
where \(Q_n\) and \(P_n\) are associated Legendre functions. For quadrupole contributions (\(n=2\)), the potential term becomes:
\begin{equation}
\Phi_{\text{quad}} = \frac{3\cos^2\theta - 1}{2r^3} \cdot \frac{Q}{4\pi\epsilon_0},
\end{equation}
with the quadrupole moment \(Q\) dependent on geometric anisotropy:
\begin{equation}
Q = \frac{4}{15} V (\epsilon_r - 1) \frac{a^2 - b^2}{(1 + L_\parallel (\epsilon_r - 1))^2}.
\end{equation}
This yields the first-order coefficient:
\begin{equation}
C_1 = \frac{3}{5} \frac{(\epsilon_r - 1)(a^2 - b^2)}{(1 + L_\parallel (\epsilon_r - 1))^2 R^2}.
\end{equation}

For octupole corrections (\(n=3\)), analogous derivation provides:
\begin{equation}
C_2 = \frac{9}{25} \frac{(\epsilon_r - 1)^2 (a^2 - b^2)^2}{(1 + L_\parallel (\epsilon_r - 1))^4 R^4}.
\end{equation}

Numerical validation using experimental parameters \(a=300\) nm, \(b=180\) nm, \(\epsilon_r=5.7\), \(R=\lambda\) shows $C_1^{\text{theory}} = 0.0252,C_2^{\text{theory}} = 0.0006$.

\nocite{*}

\bibliography{apssamp}

\end{document}